\documentclass[12pt,a4paper]{article}

\usepackage{amsmath,amssymb}

\usepackage{todonotes}
\usepackage[noend]{algorithmic}
\usepackage[boxed]{algorithm2e}
\usepackage{hyperref}

\newcommand{\nb}{\ensuremath{\mathsf{nb}}}

\renewcommand{\vec}[1]{\mathbf{#1}}

\newcommand{\LineIf}[2]{ \STATE \algorithmicif\ {#1}\ \algorithmicthen\ {#2} }

\newcommand{\mstd}{\ensuremath{\mathsf{mstd}}}
\newcommand{\mspd}{\ensuremath{\mathsf{mspd}}}
\newcommand{\can}{\ensuremath{\mathsf{can}}}
\newcommand{\cmspd}{\ensuremath{\mathtt{PD_k}}}
\newcommand{\cmstd}{\ensuremath{\mathtt{TD_k}}}

\newtheorem{theorem}{Theorem}
\newtheorem{lemma}[theorem]{Lemma}

\newtheorem{observation}[theorem]{Observation}
\newtheorem{definition}{Definition}
\newtheorem{claim}{Claim}

\newenvironment{proof}{\noindent   {\bf Proof. }}{\hspace*{\fill}$\Box$\par\vspace{2mm}}

\begin{document}

\title{Subexponential time algorithms for finding small tree and path decompositions}
\author{Hans L. Bodlaender\thanks{Utrecht University and Technical University Eindhoven, the Netherlands. Email: 
\href{mailto:H.L.Bodlaender@uu.nl}{H.L.Bodlaender@uu.nl}. Partially supported by the Networks project, funded by the Dutch Ministry of Education, Culture and Science through NWO.} \and
Jesper Nederlof\thanks{Technical University Eindhoven. Email:\href{mailto:j.nederlof@tue.nl}{j.nederlof@tue.nl}. Supported by NWO Veni project 639.021.438}}

\maketitle

\begin{abstract}
The Minimum Size Tree Decomposition (MSTD) and Minimum Size Path Decomposition (MSPD) problems ask for a given $n$-vertex graph $G$ and integer $k$, what is the minimum number of bags of a tree decomposition (respectively, path decomposition) of $G$ of width at most $k$. The problems are known to be NP-complete
for each fixed $k\geq 4$.
We present algorithms that solve both problems for fixed $k$ in $2^{O(n/ \log n)}$ time and show that they cannot be solved in $2^{o(n / \log n)}$ time, assuming the Exponential Time Hypothesis.
\end{abstract}

\section{Introduction}
In this paper, we consider two bicriteria problems concerning path and tree decompositions, namely, for an integer $k$,
find for a given graph $G$ a path or tree decomposition with the minimum number of bags. For both problems, we give
exact algorithms that use $2^{O(n / \log n)}$ time and give a matching lower bound, assuming the Exponential Time Hypothesis.
The results have a number of interesting features. To our knowledge, these are the first problems for which a matching
upper and lower bound (assuming the ETH) with the running time $2^{\Theta(n / \log n)}$ is known. The algorithmic technique
is to improve the analysis of a simple idea by van Bodlaender and van Rooij~\cite{BodlaendervR11}: a branching algorithm with memorization would use
$2^{O(n)}$ time, but combining this with the easy observation that isomorphic subgraphs give rise to equivalent subproblems, by adding isomorphism
tests on the right locations in the algorithm, the savings in time is achieved. Our lower bound proofs use a series
of reductions; the intermediate problems in the reductions seem quite useful for showing hardness for other problems.

Bicriteria problems are in many cases more difficult than problems with one criterion that must be optimized. For the problems that we consider in this paper, this is not different: if we just ask for a tree or path decomposition with the minimum number of bags, then the problem is trivial as there always is a tree or path decomposition with one bag. Also, it is well known that the problem to decide if the treewidth or pathwidth of a graph is bounded by a given number $k$ is fixed parameter tractable. However, recent results show that if we ask to minimize the number of bags of the tree or path decomposition of width at most $k$, then the problem becomes para-NP-complete (i.e., NP-complete for some fixed $k$) as shown in~\cite{DereniowskiKZ15,LiMN13}.

The problem to find path decompositions with a bound on the width and a minimum number of bags was first studied by
Dereniowski et al.~\cite{DereniowskiKZ15}. Formulated as a decision problem, the problem MSPD$_k$ is to determine, given a graph $G=(V,E)$ and integer $s$, whether $G$ has a path decomposition of width at most $k$ and with at most $s$ bags. Dereniowski et al.~\cite{DereniowskiKZ15}
mention a number of applications of this problem and study
the complexity of the problem for small values of $k$.
They show that for $k\geq 4$, the problem is NP-complete,
and for $k\geq 5$, the problem is NP-complete for connected graphs. They also give polynomial time algorithms for the
MSPD$_k$ problem for $k\leq 3$ and discuss a number of applications
of the problem, including the Partner Units problem, problems in scheduling, and in graph searching.  

Li et al.~\cite{LiMN13} introduced the MSTD$_k$ problem:
given a graph $G$ and integer $\ell$, does $G$ have a tree
decomposition of width at most $k$ and with at most $\ell$
bags. They show the problem to be NP-complete for $k\geq 4$
and for $k\geq 5$ for connected graphs, with a proof similar
to that of Dereniowski et al.~\cite{DereniowskiKZ15} for the
pathwidth case, and show that the problem can be solved
in polynomial time when $k\leq 2$.

In this paper, we look at exact algorithms for the
MSPD$_k$ and MSTD$_k$ problems. Interestingly, these
problems (for fixed values of $k$) allow for subexponential time
algorithms. The running time of our algorithm is of a form that
is not frequently seen in the field: for each fixed $k$, we give
algorithms for MSPD and MSTD that use $2^{O(n/\log n)}$ time. Moreover, we show that these results are tight in the sense that there are no $2^{o(n/\log n)}$ time algorithms for MSPD$_k$ and MSTD$_k$ for some large enough $k$, assuming the Exponential Time Hypothesis.

Our algorithmic technique is a variation and extension of 
the technique used by Bodlaender and van Rooij~\cite{BodlaendervR11} for subexponential time algorithms for
{\sc Intervalizing $k$-Colored Graphs}. That algorithm has the same running time as ours; we conjecture
a matching lower bound (assuming the ETH) for {\sc Intervalizing 6-Colored Graphs}.


\section{Preliminaries}

\paragraph{Notation}
In this paper, we interpret vectors as strings and vice versa whenever convenient, and for clarity use boldface notation for both. When $\vec{a},\vec{b} \in \Sigma^\ell$ are strings, we denote $\vec{a}||\vec{b}$ for the string obtained by concatenating $\vec{a}$ and $\vec{b}$. We let $s^{\ell}$ denote the string repeating symbol $s$ $\ell$ times. Also, we denote $\vec{a} \preceq \vec{b}$ to denote that $a_i \leq b_i$ for every $1\leq i \leq n$ and use $\vec{1}$ to denote the vector with each entry equal to $1$ (the dimension of $\vec{1}$ will always be clear from the context). We also add vectors, referring to component-wise addition.

\paragraph{Tree and Path decompositions.}
Unless stated otherwise, the graphs we consider in this paper are simple
and undirected. We let $n=|V|$ denote the number of vertices of the graph $G=(V,E)$.
A {\em path decomposition} of a graph $G=(V,E)$ is a sequence of
subsets of $V$: $(X_1, \ldots, X_s)$ such that
\begin{itemize}
\item $\bigcup_{1\leq i\leq s} X_i = V$,
\item For all edges $\{v,w\}\in E$: there is an $i$, $1\leq i\leq s$, with $v,w\in X_i$,
\item For all vertices $v\in V$: there are $i_v$, $j_v$, such that $i \in [i_v,j_v] \Leftrightarrow v\in X_i$.
\end{itemize}
The {\em width} of a path decomposition $(X_1, \ldots, X_s)$ is
$\max_{1\leq i\leq s} |X_i|-1$; its {\em size} is $s$.
The {\em pathwidth} of a graph $G$ is the minimum width of a path decomposition of $G$.
We will refer to $X_s$ as the {\em last bag} of $(X_1, \ldots, X_s)$.
A {\em tree decomposition} of a graph $G=(V,E)$ is a pair $(\{X_i~|~ i\in I\}, T=(I,F))$ with
$\{X_i ~|~ i\in I\}$ a family of subsets of $V$, and $T$ a rooted tree, such that
\begin{itemize}
\item $\bigcup_{i\in I} X_i = V$.
\item For all edges $\{v,w\}\in E$: there is an $i\in I$ with $v,w\in X_i$.
\item For all vertices $v\in V$: the set $I_v = \{i\in I~|~ v\in X_i\}$ induces a
subtree of $T$ (i.e., is connected.)
\end{itemize}
The {\em width} of a tree decomposition $(\{X_i ~|~ i\in I\},T=(I,F))$ is
$\max_{i\in I} |X_i|-1$; its {\em size} is $|I|$.
The {\em treewidth} of a graph $G$ is the minimum width of a tree decomposition of $G$.
In the definition above, we assume that $T$ is rooted: this does not change the minimum width or size, but makes proofs slightly easier. Elements of $I$ and
numbers in $\{1, 2, \ldots, s\}$ and their corresponding sets $X_i$ are called 
{\em bags}.

A tree decomposition $(\{X_i~|~ i\in I\}, T=(I,F))$ of a graph $G=(V,E)$
is {\em nice}, if for each $i\in I$, one of the following cases holds:
\begin{itemize}
\item $i$ has two children $j_1$ and $j_2$ with $X_i = X_{j_1} = X_{j_2}$ ({\em join node}.)
\item $i$ has one child $j$ and there is a $v\in V$ with $X_i = X_j \cup \{v\}$ ({\em introduce node}.)
\item $i$ has one child $j$ and there is a $v\in V$ with $X_i = X_j \setminus \{v\}$ ({\em forget node}.)
\item $i$ has no children ({\em leaf node}).
\end{itemize}

The following result is folklore.

\begin{lemma}[\cite{Kloks93}, Lemma 13.1.2]\label{lem:nicetree}
Let $G=(V,E)$ have treewidth at most $k$. Then $G$ has a nice tree decomposition
of width at most $k$ with at most $4 n$ bags. 
\end{lemma}

Usually, nice tree and path decompositions have more than the minimum number of
bags. The notions are still very useful for our analysis; in particular, they help to
count the number of non-isomorphic graphs of treewidth or pathwidth at most $k$, as we see shortly.

We will use the following fact.
A weakly binary tree is a rooted tree where each vertex has at most two children.

\begin{lemma}[Otter 1948 \cite{Otter48}]\label{lem:otter}
The number of non-isomorphic weakly binary trees with $n$ vertices is bounded by $O(2.484^n)$.
\end{lemma}

\begin{lemma}\label{lem:nonisomtw}
The number of non-isomorphic graphs with $n$ vertices of treewidth at most $k$ is 
$2^{O(kn)}$.
\end{lemma}

\begin{proof}
By Lemma~\ref{lem:nicetree}, we have a nice tree decomposition $(\{X_i\},T)$ of width at most $k$ and size
at most $4n$. 

It is well known that we can color the vertices of a graph with treewidth $k$ by $k+1$ colors,
such that all vertices in a bag have a different color. (The inductive proof is as follows:
this clearly holds if we have a tree decomposition with one bag. Otherwise, take a leaf bag $i$
with neighboring bag $j$. Inductively, color all vertices except the vertices in $X_i-X_j$;
the vertices in $X_i-X_j$ do not belong to any bag other than $X_i$; color these with colors,
different from all other vertices in $X_i$.)

To bound the number of non-isomorphic graphs of treewidth $k$ with $n$ vertices,
we associate with each such graph a nice tree decomposition of width $k$ with the vertices
colored as above, and multiply a bound on the number of underlying binary trees
by a bound on the number of non-isomorphic cases for the bags. The number of underlying
binary trees is bounded by $2^{O(n)}$, by Otter's result (Lemma~\ref{lem:otter}).

To obtain our bound, we note that for each edge $e=\{v,w\}\in E$, there is exactly one node
$i_e$ in the tree decomposition, such that $\{v,w\}\in X_{i_e}$ and 
$i_e$ is the root of the tree decomposition or the parent of $i_e$ is a forget node that
forgets $v$ or forgets $w$. This enables us to count the number of possibilities for edges
in the graph when looking at forget nodes, with the root as a simple special case.

Let us now count the number of possibilities of each non-root bag of $T$, distinguishing on its type: 
\begin{itemize}
	\item\textbf{Join node.} There is no additional information: $1$ possibility.
	\item\textbf{Introduce.} At introduce nodes, we only determine what color the newly
	introduced vertex has. This gives at most $k+1$ possibilities.
	\item\textbf{Forget.} At a forget node, we have at most $(k+1)\cdot 2^k$ possibilities:
	we identify the forgotten vertex by it color ($k+1$ possibilities), and each of the
	other at most $k$ vertices in the bag below the forget node can have an edge to the
	forgotten vertex or not ($2^k$ possibilities).
	\item\textbf{Leaf.} For each of the $k+1$ colors, there can be a vertex with this color
	in the leaf bag or not: $2^{k+1}$ possibilities.
\end{itemize}

We thus have $2^{O(k)}$ possibilities per non-root bag; we have $O(n)$ bags, which gives
an upper bound of $2^{O(kn)}$. We still need to count the number
of different possibilities concerning the edges between vertices in the root bag: for each
pair of vertices in the root bag, there can be an edge or not ---
 this
gives an extra multiplicative factor of $2^{O(k^2)}$, but as $k < n$, the number of combinations
stays $2^{O(kn)}$.
\end{proof}

\section{Path and tree decompositions with few bags}
\subsection{Finding path decompositions with memorization and isomorphism tests}
\label{section:treewidthalgorithm}
In this section, we describe our algorithm for the MSPD problem.
Throughout the section, we assume that $k$ is a fixed positive integer and that $G$ has treewidth at most $k$ (note that we can determine this in linear time for fixed $k$ (cf.~\cite{Bodlaender96}) and return NO if the treewidth is higher than $k$). Our branching algorithm is parameterized by `a good pair', formalized as follows:

\begin{definition}
A {\em good pair} is a pair of vertex sets $(X,W)$, such that
\begin{itemize}
\item $|X|\leq k+1$, 
\item $X\cap W = \emptyset$, and
\item for all
$v\in W$, $N(v)\subseteq W\cup X$. Equivalently, $W$ is the union of
the vertex sets of zero or more connected components of $G[V \setminus X]$.
\end{itemize}

For a good pair $(X,W)$, let $\mstd_k(X,W)$ ($\mspd_k(X,W)$) be the minimum $s$ such
that there is a tree (path) decomposition of $G[X\cup W]$ of width at most $k$, where $X$ is the root
bag (last bag) of the tree (path) decomposition.
\end{definition}

\paragraph{A recursive formulation for path decompositions.}
The following lemma gives a recursive formulation for $\mspd_k$. 
The formulation is the starting point for our algorithm, but we will
in addition exploit graph isomorphisms, see below.

\begin{lemma}\label{lem:pdrec}
If $|X|\leq k+1$, then $\mspd_k(X, \emptyset) = 1$. Otherwise, let $(X,W)$ be a good pair, and $W\neq \emptyset$. Then 
\begin{equation}\label{eq:pwrec}
	\mspd_k(X,W) = \min_{\substack{Y \subseteq X \cup W \\ X\neq Y \\ W \cap N(X\setminus Y)=\emptyset}}1+\mspd_k(Y,W\setminus Y).
\end{equation}
\end{lemma}

\begin{proof}
The first part with $|X|\leq k+1$ is trivial: take the only path decomposition with one bag $X$.

Otherwise, suppose $Y$ fulfills $W \cap N(X \setminus Y) =\emptyset$.
Let $PD=(X_1, \ldots, X_s)$ be a path decomposition of width at most $k$
of $G[Y \cup W]$ with $X_s=Y$.
Now we verify that $(X_1, \ldots, X_s, X)$ is a path decomposition of width at most $k$ of $G[X \cup W]$. Since there can be no edges between $X\setminus Y$ and $W$, all the edges incident to $X\setminus Y$ are covered in the bag $X$ and all other edges are covered in $PD$ since it is a path decomposition of $G[Y \cup W]$. Also, a vertex $v \in X \setminus Y$ cannot occur in $PD$ so the bags containing a particular vertex will still induce a connected part in the path decomposition.

Conversely, suppose $(X_1, \ldots, X_s,X)$ is a minimal size path decomposition of width at most $k$
of $G[X \cup W]$. Note that $X=X_s$ contradicts this path decomposition being of minimal size, so we may assume $X\neq X_s$. Vertices in $X \setminus X_s$ do not belong to $\bigcup_{1\leq i\leq s-1} X_i$, by the definition of path decomposition, so we must have that $W \cap N(X\setminus X_s)=\emptyset$ since otherwise not all edges incident to $X$ are covered in the path decomposition. Hence, $X_s$ fulfills all the conditions of the minimization in the recurrence.

We have that $(X_1, \ldots, X_s)$ is a path decomposition of $G[X_s \cup (W\setminus X_s)]$, which has at least $\mspd_k(X_{s-1},W\setminus X_{s-1})$ bags and hence taking $Y=X_s$ shows that $\mspd_k(X,W) \leq s+1$. 
\end{proof}

\paragraph{Isomorphism.}
The following notion will be needed for presenting the used recurrence for tree decompositions and essential for quickly evaluating~\eqref{eq:pwrec}. Intuitively, it indicates $G[X \cup W]$ being isomorphic to $G[Y\cup Z]$ with an isomorphism that maps $X$ to $Y$. More formally,
\begin{definition}\label{def:iso}
Good pairs $(X,W)$ and $(X,Z)$ are \emph{isomorphic} if there is a bijection $f: X\cup W \leftrightarrow X\cup Z$, such that
\begin{enumerate}
\item For all $v, w\in X\cup W$: $\{v,w\}\in E \Leftrightarrow \{f(v),f(w)\}\in E$, and
\item $f(v)=v$ for all $v\in X$.
\end{enumerate}
\end{definition}

We will use the following obvious fact:
\begin{observation}\label{obs:iso}
Suppose good pair $(X,W)$ is isomorphic to good pair $(X,Z)$. Then
$\mspd_k(X,W)=\mspd_k(X,Z)$ and $\mstd_k(X,W)=\mstd_k(X,Z)$
\end{observation}

In our algorithm we use a result by Loksthanov et al.~\cite{LokshtanovPPS14} which
gives an algorithm that for fixed $k$ maps each graph $G$ of treewidth
at most $k$ to a string $\can(G)$ (called its {\em canonical form}), such that two graphs $G$ and $H$ are isomorphic if
and only if $\can(G)=\can(H)$. The result also holds for graphs where vertices (or
edges) are labeled with labels from a finite set and the isomorphism should map vertices to vertices of the same label. We can use this result to make canonical forms for good pairs:
\begin{observation}\label{obs:iso2}
An isomorphism class of the good pairs $(X,W)$ can be described by the triple $\can(X,W):=(X,\can(G[X \cup W],f)$ where $f$ is a bijection from $X$ to $|X|$.
\end{observation}

Here, $f: X \leftrightarrow X$ can be (for example) be defined as the restriction of $\pi$ onto $X$ of the lexicographically smallest (with respect to some arbitrary ordering) isomorphism $\pi$ of $G[X \cup W]$.

\paragraph{A recursive algorithm with memorization.}
We now give a recursive algorithm \cmspd{} to compute for a given good pair
$(X,W)$ the value $\mspd_k(X,W)$. The algorithm uses memorization. In a data structure $D$, we store values that we have computed. We can use
e.g., a balanced tree or a hash table for $D$, that is initially assumed empty. Confer Algorithm~\ref{alg:find1}.

\begin{algorithm}
\label{alg:find1}
\caption{Finding a small path decompositions of width at most $k$.}
\begin{algorithmic}[1]
\REQUIRE \cmspd($X$,$W$)
	\LineIf{$|X|\leq k+1$ and $W=\emptyset$}{\algorithmicreturn\ $1$}
	\LineIf{$D(\can(X,W))$ is stored}{\algorithmicreturn\ $D(\can(X,W))$}
	\STATE $m \gets \infty$.
	\FORALL{$Y \subseteq X\cup W$ such that $Y\neq X, N(X \setminus Y)\subseteq X, |Y| \leq k$}
		\STATE $m \gets \min \{m,1+ \cmspd(Y, W \setminus Y)\}$.
	\ENDFOR
	\STATE Store $D(\can(X,W))\gets m$.
	\RETURN $m$.
\end{algorithmic}
\end{algorithm}

The correctness of this method follows directly from Lemma~\ref{lem:pdrec} and Observation~\ref{obs:iso}. The main difference with a traditional evaluation
(with memorization) of the recursive formulation of $\mspd$ is that we 
store and lookup values under their canonical form under isomorphism --- 
this simple change is essential for obtaining a subexponential running
time. The fact that we work with graphs of bounded treewidth and for
these, {\sc Graph Isomorphism} is polynomial \cite{Bodlaender90b,LokshtanovPPS14}
makes that we can perform this step sufficiently fast.

Equipped with the \cmspd{} algorithm, we solve the MSPD problem as
follows: for all $X\subseteq V$ with $|X|\leq k+1$, 
run $\cmspd(X, V\setminus X)$; report the smallest
value over all choices of $X$. 

\subsection{The number of good pairs}
We now will analyze the number of good pairs. This is
the main ingredient of the analysis of the running time of the
algorithm given above.

\begin{theorem}\label{theorem:goodpairs}
Let $k$ be a constant.
Let $G$ be a graph with $n$ vertices and treewidth at most $k$.
Then $G$ has $2^{O(n/\log n)}$ non-isomorphic good pairs.
\end{theorem}

\begin{proof}
Let us define a \emph{basic good pair} as a good pair $(X,W)$ where $G[W]$ is connected. The isomorphism classes (with respect to the notion of isomorphism from Definition~\ref{def:iso}) of good pairs can be described as follows:
let $X$ be a set of size at most $k$. Let $\mathcal{C}_1,\ldots,\mathcal{C}_\ell$ be a partition of the connected components of $G[V\setminus X]$ into basic good pair isomorphism classes, e.g.: we have for two connected components $C_a,C_b$ that $C_a,C_b \in \mathcal{C}_i$ for some $i$ if and only if there exists a bijection $X\cup C_a \leftrightarrow X\cup C_a$ such that for all $v \in X$ we have $f(v)=v$ and for all $v, w\in X\cup C_a$: $\{v,w\}\in E \Leftrightarrow \{f(v),f(w)\}\in E$. We order the isomorphism classes arbitrarily (e.g., in some lexicographical order).

Then an isomorphism class of \emph{all} good pairs can be described by a triple $(X,\vec{s}=\{c_1,\ldots,c_s\},f)$ where $c_i$ is the number of connected components of $G[V\setminus X]$ in basic pair isomorphism class $\mathcal{C}_i$. Then we have the following bound:

\begin{claim}\label{clm:smalliso}
There is a $c>0$, such that for 
a constant $k$, the number of isomorphism classes of basic good pairs $(X,W)$ with $|W|+|X|\leq \tfrac{1}{ck}\log n$ is at most $\sqrt{n}$ for sufficiently large $n$.
\end{claim}
\begin{proof}
For each $c>0$,
by Lemma~\ref{lem:nonisomtw}, the number of graph isomorphism classes of $G[X \cup W]$ 
with $|W|+|X|\leq \tfrac{1}{ck}\log n$ is at most $2^{O(\frac{k}{ck}\log n)}$ since we assumed the treewidth to be at most $k$ (as stated in the beginning of this section).

The isomorphism class of a basic good pair is described by the set $X$, the permutation of $X$ and the graph isomorphism class of $G[W \cup X]$, thus we have that the number of basic good pair isomorphism classes is at most $k!2^{\tfrac{O(kn)}{ck}\log n}\leq 2^{O(k\log(k))+\tfrac{O(kn)}{ck}\log n}$ which is $\sqrt{n}$ for large enough $n$, and proper choice of $c$ depending on the constants hidden in the $O(\cdot)$ notation.
\end{proof}

Say an isomorphism class $\mathcal{C}_i$ is \emph{small} if $(X,W) \in \mathcal{C}_i$ implies $|X|+|W|\leq \tfrac{1}{ck}\log n$ (with $c$ as given by Claim~\ref{clm:smalliso}), and it is \emph{large} otherwise. Assume $\mathcal{C}_1,\ldots,\mathcal{C}_z$ are small. By Claim~\ref{clm:smalliso}, $z$ is at most $\sqrt{n}$. Thus, since we know $c_i \leq n$, the number of possibilities of $\vec{s}$ on the small isomorphism classes is at most $n^{O(\sqrt{n})}$. For the remaining $\ell-z$ isomorphism classes of large connected components, we have that $\sum_{j=z+1}^\ell c_i\leq ck n/\log n= O(n/\log n)$. Thus, there are only $2^{O(n/\log n)}$ subsets of the large connected components that can be in $W$. Combining both bounds gives the upper bound of $2^{O(n/\log n)}$ for the number of non-isomorphic good pairs, as desired.
\end{proof}

\subsection{Analysis of the algorithm}
In this section, we analyze the running time of Algorithm~\ref{alg:find1}. First, we note that per recursive call
we have $O(n^{k+1})$ calls of the form
$\cmspd(X, V\setminus X)$. 
Observe that each call to \cmspd{} is
with a good pair as parameter, and these good pairs on Line 4 can be enumerated with linear delay. Thus, by Theorem~\ref{theorem:goodpairs}, there are 
$2^{O(n/\log n)}$ calls to \cmspd\ that
make recursive calls to \cmspd. 
Within each
single call, we have $O(n^{k+1})$ choices for a set $Y$;
computing $s=\can(X,W)$ can be done in $O(n^5)$ time (confer~\cite{LokshtanovPPS14}), and thus,
the overhead of a single recursive call is bounded by $O(n^{\max\{k+1,5\}})$.
Putting all ingredients together shows that the algorithm
uses $2^{O(n/\log n)}$ time.

\begin{theorem}
For fixed $k$, the MSPD problem can be solved in $2^{O(n/\log n)}$ time.
\end{theorem}

\subsection{Extension to finding tree decompositions}
Now we discuss how to extend the algorithm for solving the MSTD problem. Note that, like usual, dealing with tree decompositions instead of path decompositions amounts to dealing with join bags. We have the following analogue of Lemma~\ref{lem:pdrec}.
\begin{lemma}\label{lem:tdrec}
If $|X|\leq k+1$, then $\mstd_k(X, \emptyset) = 1$. Otherwise, let $(X,W)$ be a good pair, and $W\neq \emptyset$. Then $\mstd_k(X,W)=\min\{\mathsf{extend},\mathsf{branch}\}$ where 
\begin{equation}\label{eq:mstd}
\begin{aligned}
	 \mathsf{extend} &= \min_{\substack{Y \subseteq X \cup W \\ X\neq Y \\ W \cap N(X\setminus Y)=\emptyset}}1+\mstd_k(Y,W\setminus Y).\\
	 \mathsf{branch} &= \min_{\substack{W_1 \subseteq W\\ N(W_1)\subseteq W_1 \cup X}} \mstd(X,W_1) + \mstd(X,W\setminus W_1)-1.
\end{aligned}
\end{equation}
\end{lemma}
\begin{proof}
The cases $\mathsf{extend}$ and $\mathsf{branch}$ refer to whether the root bag $r$ with vertex set $X$ has exactly one child, or at least two children.
If $r$ has one child, then the same arguments that show~\eqref{eq:pwrec} can be used to show correctness of the $\mathsf{extend}$ case.
If $r$ has two or more children, then we can guess the set of vertices $W_1\subseteq W$ that appear in bags in the subtree rooted by the first child
of $r$. We must have that $W_1$ is a union of connected components of $G[W]$ by the definition of tree decompositions. Thus, the tree
decomposition can be obtained by taking a tree decomposition of $G[X\cup W_1]$ and a tree decomposition
of $G[X \cup (W\setminus W_1)]$, both with $X$ as the vertex set of the root bag, and then taking the union, identifying the two root bags.
The number of bags thus equals the minimum number of bags for the first tree decomposition (which equals $\mstd(X,W_1)$), plus the
minimum number of bags for the second (equally $\mstd(X,W\setminus W_1)$), subtracting one as we counted the bag with vertex set $X$ twice.
\end{proof}

Given Algorithm~\ref{alg:find1} and~\eqref{eq:mstd}, the algorithm for computing \mstd\ suggests itself since it is easy to see that again we only need to evaluate $\mstd_k(X,W)$ for good pairs $(X,W)$. This is indeed our approach but there is one small complication, since we cannot compute $\mathsf{branch}$ in a naive way because the number of connected components of $G[W]$ could be $\Omega(n)$. We deal with this by even further restricting the set of subsets of $W$ we iterate over, again based on Observation~\ref{obs:iso}.

\begin{algorithm}
\label{alg:find2}
\caption{Extension of Algorithm~\ref{alg:find1} to find small tree decompositions of width at most $k$.}
\begin{algorithmic}[1]
\REQUIRE \cmstd($X$,$W$)
	\LineIf{$|X|\leq k+1$ and $W=\emptyset$}{\algorithmicreturn\ $1$}
	\LineIf{$D(\can(X,W))$ is stored}{\algorithmicreturn\ $D(\can(X,W))$}
	\STATE $m \gets \infty$
	\FORALL{$Y \subseteq X\cup W$ such that $Y\neq X$ \algorithmicand\ $N(X \setminus Y)\subseteq X$}
		\STATE $m \gets \min \{m,1+ \cmstd(Y, W \setminus Y)\}$
	\ENDFOR
	\STATE Let $\mathcal{C}_1,\ldots,\mathcal{C}_\ell$ be the isomorphism classes of the basic good pairs $(X,W')$,\\ where $W'$ is a connected component of $W$
	\STATE For $1\leq i\leq \ell$, let $c_i$ be the number of $(X,W') \in \mathcal{C}_i$ where $W'$ is a connected component of $W$
	\STATE For $1\leq i\leq \ell$ and $0\leq j \leq c_i$, let $W^i_j$ be the union of the $j$ lexicographically\\ first connected components $W'$ such that $(X,W') \in \mathcal{C}_i$
	\FORALL{vectors $\vec{y} \preceq (c_1,\ldots,c_\ell)$}
		\STATE  $W_1 \gets \bigcup_{i=1}^\ell W^i_{y_i}$
		\STATE  $W_2 \gets W \setminus W_1$
		\STATE  $m \gets \min\{m, \cmstd(X,W_1) + \cmstd(X,W_2)-1\}$
	\ENDFOR
	\STATE Store $D(\can(X,W))\gets m$
	\RETURN $m$.
\end{algorithmic}
\end{algorithm}

We solve the \mstd\ problem in Algorithm~\ref{alg:find2}. Let us first discuss the correctness of this algorithm. Note that similarly as in Algorithm~\ref{alg:find1}, it implements the memorization with the datastructure $D$. It is easy to see that after Line~5, $m$ equals the quantity $\mathsf{extend}$ from~\eqref{eq:mstd}. By Lemma~\ref{lem:tdrec}, it remains to show that at Line~13, $m$ equals $\min\{\mathsf{extend},\mathsf{branch}\}$.

To see this, note that by construction we iterate over a subset of $2^{W}$ generating all isomorphism classes that $(X,W_1)$ subject to $N(W_1)\subseteq W_1 \cup X$ can generate, and by Observation~\ref{obs:iso} this is sufficient to find any optimal partition of $W$ into $W_1,W_2$.

%
%
%
%

\section{Lower bounds}
This section is devoted to the proof of the following theorem:

\begin{theorem}
Suppose the Exponential Time Hypothesis holds, then there is no algorithm for
MSPD or MSTD for any fixed $k\geq 39$ using $2^{o(n / \log n)}$ time.
\label{theorem:mspdlow}
\end{theorem}

\subsection{Lower bounds for intermediate problems}

In order to obtain Theorem~\ref{theorem:mspdlow}, we derive a number
of intermediate results, some of which are of independent interest, as they
can be used as good starting points for other lower bound results.
Thus, we show for a number of problems that they have
no subexponential time algorithm, assuming that the Exponential Time Hypothesis holds.
We start with the following problem.

\begin{verse}
{\sc Partition Into Triangles for 4-regular 3-colorable graphs}\\
{\bf Given:} Graph $G=(V,E)$ with a proper vertex coloring $c: V \rightarrow \{1,2,3\}$,
such that each vertex in $G$ has degree four and each vertex belongs to at most three triangles in $G$.
\\
{\bf Question:} Can we partition to vertices in $n/3$ sets $V_1, \ldots, V_{n/3}$ of three elements each, 
such that each set forms a triangle in $G$?
\end{verse}

We build upon a result by van Rooij et al.~\cite{vanRooijKB12}, who consider the more general
case of 4-regular graphs, and show that for this version that there is no subexponential time
algorithm assuming the ETH. We modify their proof to obtain the following result.

\begin{theorem}
Suppose that there is an algorithm for {\sc Partition Into Triangles for 4-regular 3-colorable graphs}
that uses $2^{o(n)}$ time. Then the Exponential Time Hypothesis does not hold.
\label{theorem:triangles}
\end{theorem}

\begin{proof}
van Rooij et al.~\cite{vanRooijKB12} observe that a construction by Schaefer \cite{Schaefer78b} gives that there is
no subexponential time algorithm for {\sc Exact 3-Satisfiability} unless the Exponential Time Hypothesis
does not hold. 
{\sc Exact 3-Satisfiability} is the version of {\sc 3-Satisfiability}, where we require that each clause
contains exactly one true literal. Using this observation and the Sparsification Lemma~\cite{ImpagliazzoPZ01}, we can assume that we have an
instance of {\sc Exact 3-Satisfiability} with $O(n)$ clauses. 

Given such an instance, we build an instance of {\sc Partition Into Triangles for 4-regular 3-colorable graphs}
using a number of steps. First we will define clause gadgets named fan gadgets and variable gadgets named cloud gadgets.

\paragraph{Fan gadgets.}
We take for each clause three fan gadgets. The fan gadget was used by
van Rooij et al.~\cite{vanRooijKB12}); our change here is that we use three such gadgets
instead of one. The fan gadgets and the coloring of the fan gadgets is shown in
Figure~\ref{figure:fan}. 

\begin{figure}
\begin{center}
\includegraphics[scale=0.8]{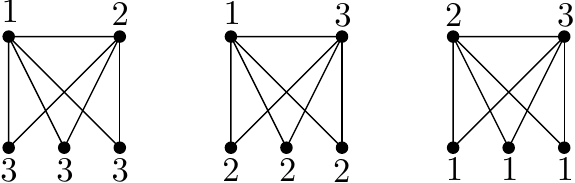}
\end{center}
\caption{A fan gadget}
\label{figure:fan}
\end{figure}

The vertices of degree two in a fan gadget represent the occurrences of literals in 
clauses. Each such occurrence is represented by three vertices, one of each color.

So for each literal $x_i$ or $\overline{x_i}$, there is an equal number of vertices
with color 1, with color 2, and with color 3 that represent this literal.

\paragraph{Cloud gadgets.}
We now build a slightly modified version of the clouds from \cite{vanRooijKB12}.
A cloud consists of a number of triangles; a cloud has vertices of degree two and
four and exactly two ways to cover the vertices of degree four by disjoint triangles;
these covers can use some vertices of degree two and leave other 
vertices of degree two untouched. In one of the covers, the positive literals
are untouched, and in the other, the negative literals are untouched.
Vertices of degree two in a cloud are labelled P or N.

We define an $(i,j)$-cloud as follows, for $i\geq 0$ and $j\geq 0$.
\begin{itemize}
	\item A $(1,0)$-cloud and a $(0,1)$ cloud is a triangle as given in Figure~\ref{figure:basicclouds}. In a $(1,0)$-cloud, all three vertices
	are labelled P, in a $(0,1)$-cloud, all three vertices are labelled N.
	\item A $(1,1)$-cloud is as given in Figure~\ref{figure:basicclouds}; the structure
is known as the star of David. We alternatingly label the vertices of degree 2 P and N.
	\item For $i\geq 1$, we build an $(i+1,j)$-cloud from an $(i,j)$ cloud
as follows: take one vertex with label P, and add the construction as shown in Figure~\ref{figure:cloudtransform}.
All new vertices of degree two are labelled P.
	\item For $j\geq 1$, we use the same construction to build a $(i,j+1)$-cloud from
a $(i,j)$-cloud, but start with a vertex with label N, and label the new vertices of degree two with N.
\end{itemize}

Note that an $(i,j)$-cloud is also a $(j,i)$-cloud.

\begin{figure}
\begin{center}
\includegraphics[scale=0.6]{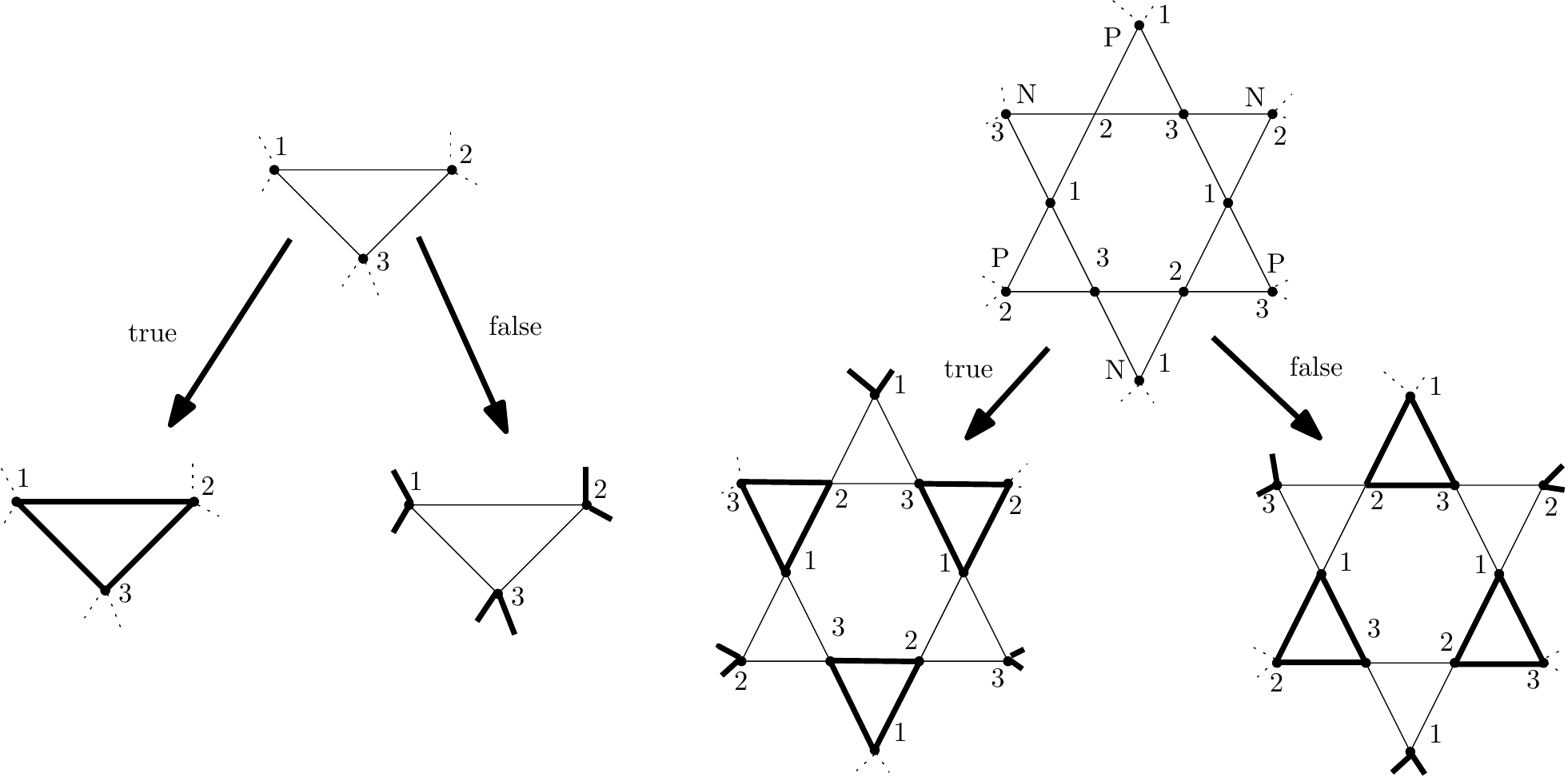}
\end{center}
\caption{Basic variable gadgets: triangle and $(1,1)$-cloud, along with how partial triangle partitions corresponding with assigning the variable to $\mathsf{true}$ and $\mathsf{false}$ look like. The numbers in the figure represent the 3-coloring.}
\label{figure:basicclouds}
\end{figure}

\begin{figure}
\begin{center}
\includegraphics[scale=0.6]{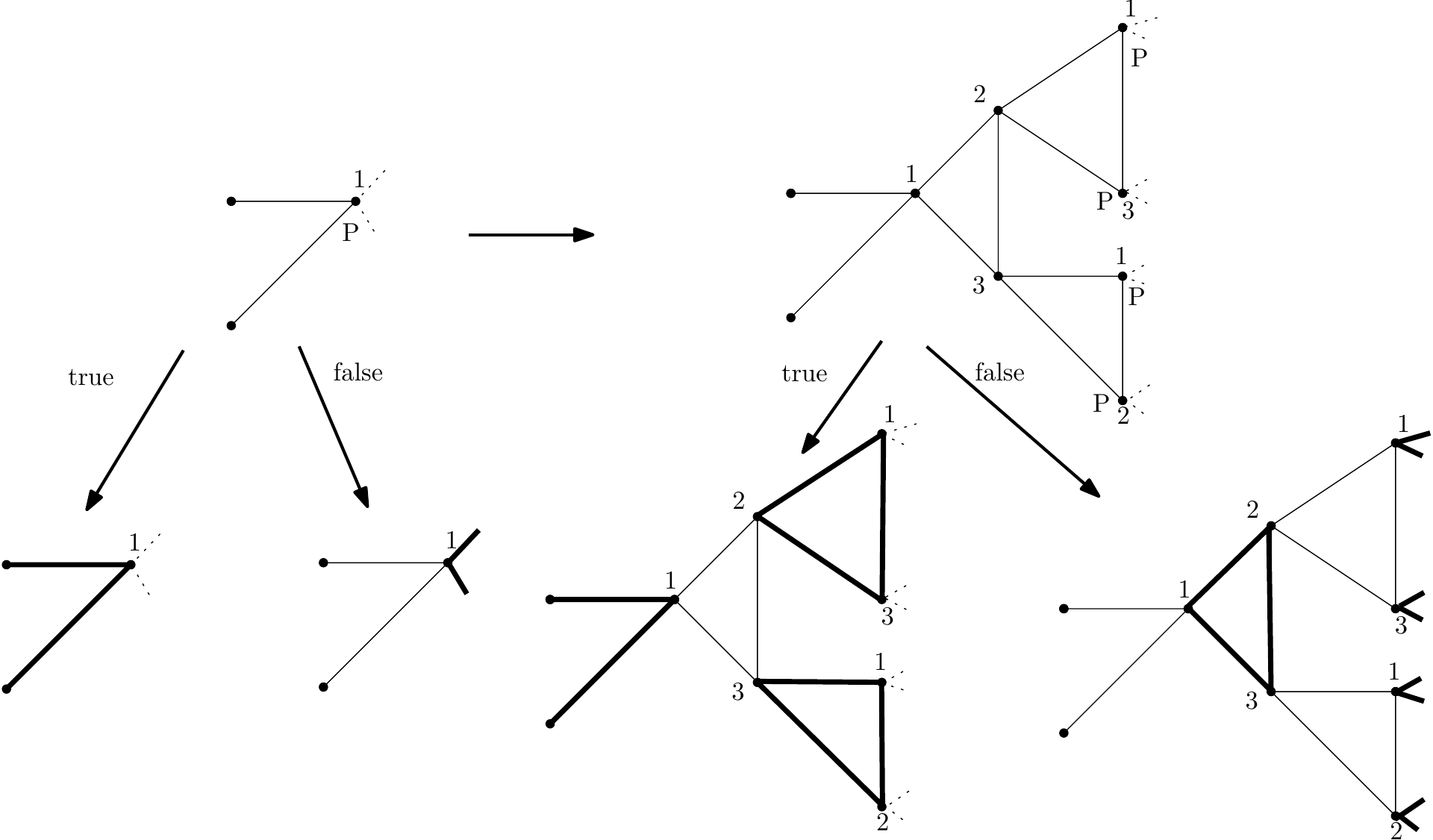}
\end{center}
\caption{Extended variable gadget to build an $(i+1,j)$-cloud from an $(i,j)$-cloud, along with how partial triangle partitions corresponding with assigning the variable to $\mathsf{true}$ and $\mathsf{false}$ look like. The numbers in the figure represent the 3-coloring.}
\label{figure:cloudtransform}
\end{figure}

The construction has the following properties. For each color in $\{1,2,3\}$,
the number of vertices with degree two with that color in the cloud increases
by exactly one; for the vertex that is replaced, we have now two vertices with that
color instead of one. We also have that an $(i,j)$-cloud has $3i$ vertices of degree 2 labelled $P$
and $3j$ vertices of degree two labelled N.
Moreover, each cloud has exactly two ways of covering all
vertices of degree four with triangles, that will correspond to setting the associated variable to true or false as depicted in Figure~\ref{figure:basicclouds} and~\ref{figure:cloudtransform}. In one cover, all $3i$ vertices with label P are used,
and in the other cover, all $3j$ vertices with label N are used; the other vertices of degree two are unused.

Based on the these possible covering, let us distinguish two types of vertices with two neighbors inside a cloud, depending on in which of the two covering they are. We see that from the $3i$ vertices of the first type, we have $i$ of color 1, $i$ of color 2, and $i$ of color 3. Similarly, the other type has $j$ vertices per color. We now completed the description of the construction of $(i,j)$-clouds, and continue with the construction of our transformation.

For each variable $x$, we take one cloud, as follows. Suppose
$x$ appears $i$ times in its positive form in the formula,
and $j$ times in its negative form $\overline{x}$. Then we construct an $(j,i)$ cloud representing $x$. Each vertex of degree two in this
cloud is identified with a vertex of degree two in a fan, in the following
way. Partition the vertices of the $(i,j)$-cloud with exactly two neighbors inside the $(i,j)$-cloud in vertices of type 1 and type 2 such that there are $i$ vertices of type 1 and $j$ vertices of type 2. Every vertex of type $1$ is then identified with a vertex in a fan that represents
a negative occurrence of the variable --- a vertex is always identified with
a vertex of the same color. A vertex of type 2 is identified with a vertex
in a fan that represents a positive occurrence of the variable in a clause,
again with the same color. By construction, we have precisely the correct
number of vertices for each color and literal.
Note that by identification, we do not create new triangles, as there
are no edges between degree two vertices in fans.

This completes the construction. It is easy to check that the
resulting graph indeed is 4-regular, has a proper vertex coloring,
and that each vertex belongs to at most three triangles.

We claim that there is a partition into triangles of the resulting graph
$G$, if and only if the instance of {\sc Exact 3-Satisfiability} has
a solution. The proof of this fact is identical to the
proof by van Rooij et al.~\cite{vanRooijKB12}. We sketch the idea
and refer to van Rooij et al.~\cite{vanRooijKB12} for more details.

For each fan, we need to take exactly one of the three triangles. One literal
in the triangle is set to true, and the other two literals are set to false.
Thus, we need to cover a cloud such that all vertices corresponding to
false literals are covered in
the cloud, and no vertices corresponding to true literals are covered in
the cloud. The construction of clouds is done precisely such that
either all true or all false literal vertices are used, and none of the other
type. 

We finally notice that the construction is linear, i.e., if we start
with a formula with $n$ variables and $m$ clauses, then the resulting
graph will have $O(n+m)$ vertices; e.g., one easily can show that $G$ has less
than $13m$ vertices. Thus, an algorithm for 
{\sc Partition Into Triangles for 4-regular 3-colorable graphs} that uses
subexponential time implies an algorithm for {\sc Exact 3-Satisfiability}
for instances with $O(n)$ clauses.  This shows the result of this theorem. 
\end{proof}

As a direct corollary, we obtain hardness for the following sparse variant
of the well known {\sc 3-Dimensional Matching} problem.

\begin{verse}
{\sc 3-Dimensional Matching with at most three triples per element}\\
{\bf Given:} Disjoint sets $A$, $B$, $C$, with $|A|=|B|=|C|$, set of
triples $T \subseteq A\times B\times C$, such that each element
from $A$, $B$, and $C$ appears in at most three triples from $T$.\\
{\bf Question:} Is there a subset $S\subseteq T$ of $n$ triples such that each 
element in $A\cup B\cup C$ appears in exactly one triple in $S$?
\end{verse}

\begin{theorem}
{\sc 3-Dimensional Matching with at most three triples per element}
has no subexponential time algorithm unless the Exponential Time Hypothesis
does not hold.
\end{theorem}

\begin{proof}
This follows directly from Theorem~\ref{theorem:triangles}: let $A$ be the vertices
of color 1, $B$ the vertices of color 2, $C$ the vertices of color $3$,
and $T$ the collection of triangles in $G$, and we obtain an equivalent
instance of {\sc 3-Dimensional Matching with at most three triples per element}.
\end{proof}

Now we consider the {\sc String 3-Groups}. Recall it is defined as follows:
\begin{verse}
{\sc String 3-Groups}\\
{\bf Given:} Sets $A,B,C \subseteq \{0,1\}^{6 \lceil \log n \rceil +1}$, with $|A|=|B|=|C|=n$\\
{\bf Question:} Choose $n$ elements from $A\times B \times C$, such that
each element in $A$, $B$, and $C$ appears exactly once in a triple, and
if $(\vec{a},\vec{b},\vec{c})$ is a chosen triple, then $\vec{a}+\vec{b}+\vec{c}\preceq \vec{1}$.
\end{verse}

%
%
%
\begin{theorem}
Suppose the Exponential Time Hypothesis holds. Then there is no algorithm for
{\sc String 3-Groups} using $2^{o(n)}$ time.
\label{theorem:groups}
\end{theorem}

\begin{proof}
Take an instance of {\sc 3-Dimensional Matching with at most three triples per element}. Suppose we have a set of triples $T \subseteq P \times Q \times R$.

Number the elements in $A$, $B$, and $C$ from $0$ to $n-1$. 
Let for $x\in A \cup B \cup C$, $\nb(x)$ be the binary representation of
the number of $x$ with $\lceil(\log n)\rceil$ bits. Let $\nb'(x) = \nb(x)||\overline{\nb(x)}$.
Note that each string $\nb'(x)$ has exactly $2 \lceil(\log n)\rceil$ bits, of
which exactly $\lceil(\log n)\rceil$ bits are 0 and $\lceil(\log n)\rceil$ bits are 1.

Write $\alpha = 2\lceil(\log n)\rceil$.

Transform this set as follows to the following collections of strings:

\paragraph{Type T} 
For each $(p,q,r)\in T$, we add to $C$ a string of the form
\[ \overline{\nb'(p)} || \overline{\nb'(q)} || \overline{\nb'(r)} || 00 \]

\paragraph{Type A} 
For each element $p\in P$, we add to $A$ a string of the form
\[ \nb'(p) || 0^\alpha || 0^\alpha || 10\]

\paragraph{Type B} 
For each element $q\in Q$, we add to $B$ a string of the form
\[   0^\alpha || \nb'(q) || 0^\alpha || 01 \]

\paragraph{Type CA and CB}
For each element $r\in R$, suppose there are $c_r$ triples in $T$
of the form $(\ast, \ast, r)$, i.e., with $r$ as third value.
Now, we add $c_r -1$ identical strings to $A$ of the form
\[ 0^\alpha || 0^\alpha || \nb'(r) || 01 \]
We also add $c_r -1$ identical strings to $B$ of the form
\[ 0^\alpha || 0^\alpha || 0^\alpha || 10 \]
The former are said to be of type CA, and the latter of type CB.

\begin{claim}
The collection of strings is a positive instance of {\sc Strings 3-Groups},
if and only if $T$ is a positive instance of {\sc 3-Dimensional Matching}.
\end{claim}

\begin{proof}
Suppose that $T$ is a positive instance of {\sc 3-Dimensional Matching}.
Suppose $S\subseteq T$ is a set of $n$ triples in $T$ that cover all elements in $P\cup Q\cup R$.
Now, group the strings as follows:
\begin{itemize}
\item For each $(p,q,r)\in S$, take a group consisting of the type T string corresponding to this triple,
the type $A$ string corresponding to $p$, and the type $B$ string corresponding to $q$.
\item For each $(p,q,r)\in T\setminus S$, we take a group consisting of the type T string corresponding to this triple,
and a CA string corresponding to $r$ and a type CB string corresponding to $r$.
\end{itemize}
One easily checks that this grouping fulfils the conditions.

Now, suppose that we can group the collection of strings is a positive instance of
{\sc Strings 3-Groups}. We build a solution to {\sc 3-Dimensional Matching} as follows.
If we have a group containing a type T element, a type A element, and a type B element,
then we put the triple corresponding to the type T element in the solution set $S$.
We claim that the resulting set $S$ of triples is a solution for the instance
of {\sc 3-Dimensional Matching}. 

First note that type A strings cannot be in a group with a type CB strings,
as both have a 1 on the one but last position, and similarly, as type B and
type CA strings end on a 1, they cannot be together in a group. Hence we
must have $n$ groups with a type T, a type A, and a type B string.
If such group has strings corresponding to $(p,q,r)$, $p'\in P$, and $q'\in Q$,
then we must have that $p=p'$ and $q=q'$. The first string starts
with $\overline{\nb'(p)}$, and the second with $\nb'(p')$. 
The construction of strings $\nb(p)$ and $\nb'(p')$ ensures that each has
an equal number of 0's and 1's, so if $p\neq p'$, then
there is a coordinate where both $\overline{\nb'(p)}$ and $\nb'(p')$ have
a 1. So $p=p'$. A similar argument shows that $q=q'$.
It thus follows that each element in $P \cup Q$ is covered by exactly one
triple in $S$.

Now consider an element $r\in R$. Each CA type string corresponding to $r$
must be in a group with a type T string, say corresponding to $(p,q,r')$.
If $r\neq r'$, we obtain a contradiction, with an argument similar as above.
Thus, if $r$ appears in $c_r$ triples in $T$, exactly $c_r-1$ of these
are in groups that contain CA type strings, and thus there is exactly
one triple in $S$ that covers $r$.
\end{proof}

We now can conclude Theorem~\ref{theorem:groups}; note that
as each element appears in at most three triples, we have that 
the number of strings in $A$, $B$, and $C$ is bounded by $3n$.
\end{proof}

Theorem~\ref{theorem:groups} forms the last preliminary step towards our main result,
Theorem~\ref{theorem:mspdlow}.

\subsection{Lower bounds for MSPD and MSTD}
We now are ready to show our main lower bound result, i.e., that, assuming the ETH, there
is no algorithm for MSPD or MSTD for $k\geq 39$ with running time $2^{o(n/\log n)}$.
We transform from an instance of {\sc String 3-Groups}, but first define some notions
that are used in our proofs.

\paragraph{Vector gadgets.}
We will use the following notions extensively:
\begin{definition}
The \emph{fingerprint} of a path decomposition $(X_1,\ldots,X_r)$ is the vector $(|X_1|,\ldots,|X_r|)$. A path decomposition is \emph{minimal} if (i) for all path decompositions $(X'_1,\ldots,X'_{r'})$ of $G$ we have $r' > r$ or if $r'=r$, then  $(|X_1|,\ldots,|X_r|) \preceq (|X'_1|,\ldots,|X'_r|)$. Graph $G$ \emph{$k$-implements} $\vec{w}\in \mathbb{N}^\ell_{>0}$ if $(i)$ every tree decomposition of $G$ of size $r$ and width $k+1$ is a path decomposition,  $(ii)$ all minimal path decompositions of size $r$ have fingerprint $\vec{w}$.
\end{definition}

A \emph{palindrome} is a vector $\vec{w}\in\mathbb{N}^r$ such that $(w_1,\ldots,w_r)=(w_r,\ldots,w_1)$. The most important part of our reduction is the gadget summarized by the following lemma:

\begin{lemma}\label{lem:gadget}
For every integer $k \geq 3$ and palindrome $\vec{w}\in \mathbb{N}^r_{>0}$ such that $\lceil 2k/3\rceil < w_i \leq k$ for all $i\leq r$, we can in polynomial time construct a graph $G$ that $k$-implements $\vec{w}$.
\end{lemma}
\begin{proof}
Construct $G$ as follows:
\begin{itemize}
	\item Construct disjoint cliques $C_0,\ldots,C_{r}$ all of size $\lfloor k/3 \rfloor$ and for $i=1,\ldots,r$ make all vertices from $C_{i-1}$ and $C_{i}$ adjacent,
	\item Construct disjoint cliques $C^p_1,\ldots,C^p_r$ where $|C^p_i|=w_i-2\lfloor k/3\rfloor$ for all $i=1,\ldots,r$ and for all $i$, make all vertices of $C^p_i$ adjacent with all vertices of $C_{i-1}$ and $C_i$.
\end{itemize}

For $i=1,\ldots,r$, let us denote $M_i=C_{i-1} \cup C_{i} \cup C^p_{i}$ for the maximal cliques of $G$. 
Since any clique must be contained in a bag of any tree decomposition we have that for every $i=1,\ldots,r$ some bag must contain a $M_i$.
Since all bags must be of width at most $k$, the maximal cliques of $G$ are of size $w_i$ for some $i$ and the maximal cliques intersect in only $\lfloor k/3\rfloor$ vertices, one bag cannot contain two maximal cliques. Hence in a path decomposition of width at most $k$ and size $r$ each bag contains exactly one maximal clique. 
Let $(\{X_i\},T)$ be a tree decomposition of width at most $k$ and size at most $r$, and suppose that $X_i$ is the bag containing $M_i$. Note that in $T$, bags $X_i$ and $X_{i+1}$ must be adjacent since they are the only bags that can contain all of $C_{i}$. 
Therefore, we know that $T$ must be a path $X_1,\ldots,X_r$ or the path $X_r,\ldots,X_1$. 
Also notice that using such a $T$ and setting $X_i=M_i$ gives us two valid tree decompositions that are path decompositions and both have $\vec{w}$ as fingerprint since $\vec{w}$ is a palindrome. Also, these are the only minimal ones since $X_i$ must contain $M_i$.
\end{proof}

\paragraph{Construction.}
Let $A,B,C$ be an instance of \textsc{String 3-Groups}. Note that without loss of generality, we way assume that all elements of $A,B,C$ are palindromes: if we change all strings $\vec{x} \in A\cup B \cup C$ to $\vec{x}||\stackrel{\leftarrow}{\vec{x}}$, where $\stackrel{\leftarrow}{\vec{x}}$ denotes the reverse of $\vec{x}$, we obtain a clearly equivalent instance where all strings are palindromes. Also, by padding zero's we may assume that for the length $\ell$ of all vectors, we have $\ell=12\lceil\log n\rceil+2$.

Let us now construct a graph $G$ such that $G$ has no tree decomposition with maximum bag size $k=53$ and size $s=n(\ell+1)$ if $(A,B,C)$ is a no-instance of \textsc{String 3-Groups}, and $G$ has a path decomposition with maximum bag size $k$ and size $s$ otherwise.

Let us denote $A=\{\vec{a^1},\ldots,\vec{a^n}\},B=\{\vec{b^1},\ldots,\vec{b^n}\},C=\{\vec{c^1},\ldots,\vec{c^n}\}$ for the binary strings in $A$, $B$, $C$. Set $k=53$, $\ell= (n-1)+ 6 n \log n$, and construct $G$ as follows
\begin{enumerate}
	\item Add one graph $G(A)$ $40$-implementing $\vec{a^1}+\vec{27}||40||\vec{a^2}+\vec{27}||40||\ldots||40||\vec{a^n}+\vec{27}||40$,
	\item For every $\vec{b^i}\in B$, add a graph $G(\vec{b^i})$ that $13$-implements $\vec{b^i}+\vec{9}$,
	\item For every $\vec{c^i}\in C$, add a graph $G(\vec{c^i})$ that $4$-implements $\vec{c^i}+\vec{3}$.
\end{enumerate}
Applying Lemma~\ref{lem:gadget} we see that all graphs $G(A)$, $G(\vec{b})$ and $G(\vec{c})$ exist and can be found in polynomial time since respectively $27 > \tfrac{2}{3}40$, $9 > \tfrac{2}{3}13$, $3 > \tfrac{2}{3}4$.

\begin{figure}
\begin{center}
\includegraphics[scale=0.68]{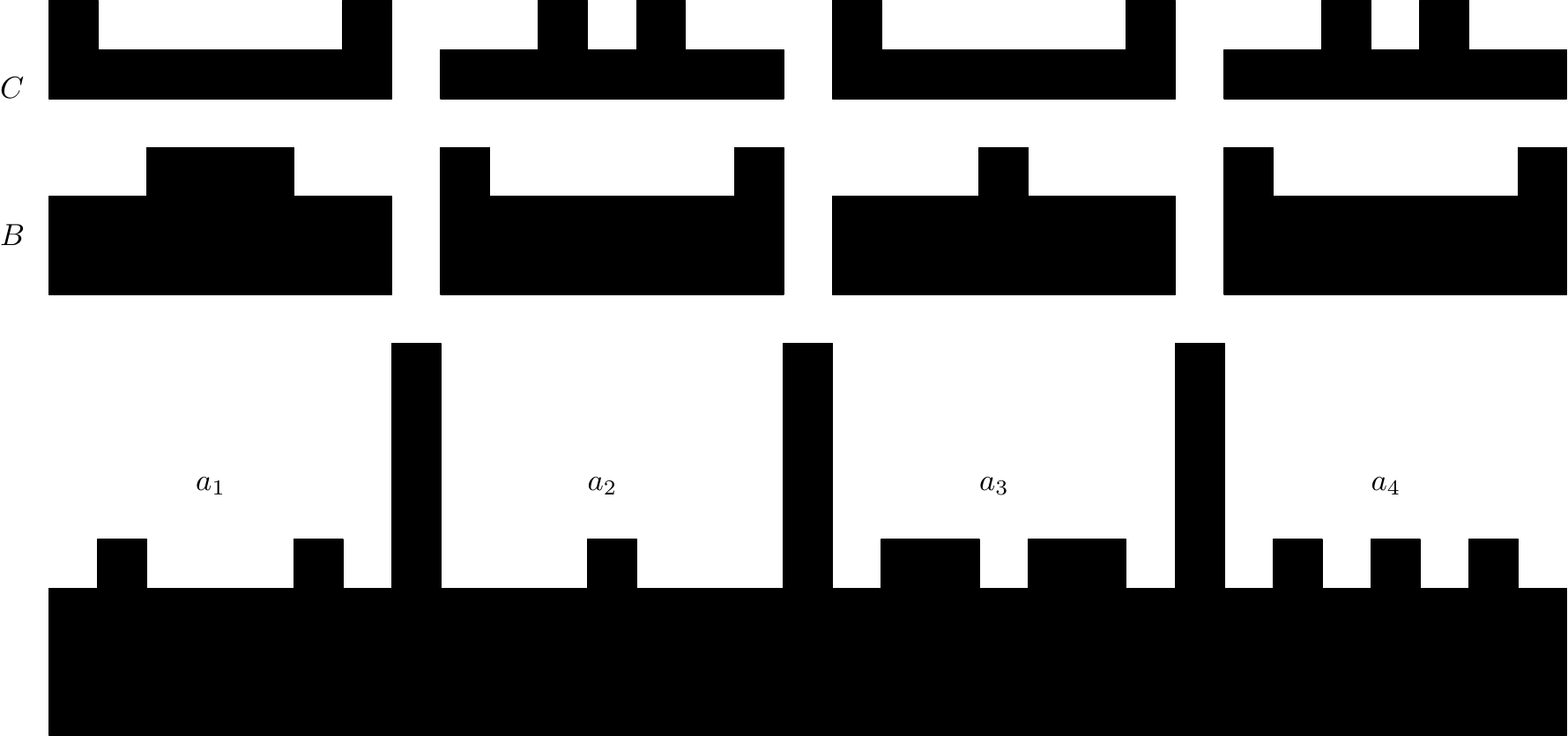}
\end{center}
\caption{Schematic illustration for the 
proof of Theorem~\ref{theorem:mspdlow}. The larger object represents all elements in $A$;
the smaller objects each represent one element from $B$ or $C$. In each `gap' between two towers, we must fit an element from $B$
and an element from $C$; $\vec{b^{i_1}}$ and $\vec{c^{i_2}}$ fit in the gap with $\vec{a^{i_3}}$, iff $\vec{a^{i_3}} + \vec{b^{i_1}} +
\vec{c^{i_2}} \preceq \vec{1}$ --- the one by one protruding blocks each represent one vertex, and we can fit at most one such vertex in the
respective bag.}
\label{figure:proofidea}
\end{figure}

Figure~\ref{figure:proofidea} gives a schematic intuitive illustration of the construction, and its correctness.

Suppose that the instance of \textsc{String 3-Groups} is a yes-instance and without loss of generality assume that $\vec{a^i}+\vec{b^i}+\vec{c^i}\preceq \vec{1}$ for all $1\leq i\leq n$. Let $(A_1,\ldots,A_{s})$ be a minimal path decomposition of $G(A)$, for $i=1,\ldots,n$ let $(B^i_1,\ldots,B^i_{\ell})$ be a minimal path decompositions of $G(\vec{b_i})$ and $(C^i_1,\ldots,C^i_{\ell})$ be a minimal path decompositions of $G(\vec{c_i})$. Then it is easy to see that 
\[
	(A_1 \cup B^1_1 \cup C^1_1,\ldots, A_\ell \cup B^1_\ell \cup C^1_\ell,A_{\ell+1}, A_{\ell+2} \cup  B^2_1 \cup C^2_1,\ldots,A_{s-1} \cup B^n_\ell \cup C^n_\ell,A_{s}),
\]
is a valid path decomposition of $G$ of size $s$. Moreover, all bags have size at most $40$: for $j$ being a positive multiple of $(\ell+1)$ we have $|A_{j}|=40$ and otherwise if $j=g(\ell+1)+i$ for $1\leq i \leq \ell+1$ then the size of the $j$'th bag equals $39+a^g_i+b^g_i+c^g_i$ which is at most $40$ by the assumption $\vec{a^i}+\vec{b^i}+\vec{c^i}\preceq \vec{1}$ for all $1\leq i\leq n$.

Suppose that $G$ has a tree decomposition $\mathbb{T}$ of width at most $k$ and size $s$. Restricted to the vertices of $G(A)$ we see that by the construction of $G(A)$, $\mathbb{T}$ has to be a path decomposition $P_1,\ldots,P_s$ where $A_i \subseteq P_i$ for all $i$ or $A_i \subseteq P_{s-i}$. These cases are effectively the same, so let us assume the first case holds. We have that there are $n$ sets of $\ell$ consecutive bags that are of size $12$ or $13$, separated with bags of size $40$. 

Then, for $t$ being a positive multiple of $\ell+1$ we have that $P_{t} \cap B^{j}_{i} = \emptyset$ for any $i,j$, and therefore for each $j$, the bags of $\mathbb{T}$ containing elements of $G(\vec{b^j})$ must be a consecutive interval of length at most $\ell$. Moreover, since all bags of $\mathbb{T}$ contain at least $27$ vertices from $G(A)$, we see that the partial path decomposition induced by $G(\vec{b})$ is of size at most $\ell$ and width at most $13$ and hence by construction it must have fingerprint $\vec{b}$. Since we have $n$ intervals of consecutive bags in $G(A)$ and $n$ graphs $G(b)$ and no two graphs can be put into the same interval we see that we can reorder $B=\{\vec{b^1},\ldots,\vec{b^n}\}$ such that if $(B^j_1,\ldots,B^j_\ell)$ is a minimal path decomposition of $G(\vec{b^j})$ then either $B^j_i \subseteq P_{j(\ell+1)+i}$ for each $i$ or $B^j_i \subseteq P_{j(\ell+1)+1-i}$ for each $i$. Note that in both cases the fingerprint of $\mathbb{T}$ induced by the vertices from $G(A)$ and $G(b)$ for each $b$ is the same.

Focusing on the vertices from $G(c)$, we have that since all bags of $\mathbb{T}$ contain at least $36$ vertices of $G(A)$ and $G(b)$ for some $b$ that the path decomposition of $G(c)$ must be of width at most $4$, and by construction thus of length at least $\ell$. By similar arguments as in the preceding paragraph, we see we may assume that $C=\{\vec{c^1},\ldots,\vec{c^n}\}$ such that either $C^j_i \subseteq P_{j(\ell+1)+i}$ for each $i$ or $C^j_i \subseteq P_{j(\ell+1)+1-i}$ for each $i$.

By the definitions of $G(A)$, $G(b)$, $G(c)$ and the assumption that $\mathbb{T}$ has width at most $53$ we then see that $\vec{a^i}+\vec{b^i}+\vec{c^i}\preceq 1$ for every $1\leq i\leq n$, as desired.

For the efficiency of the reduction: notice that the graph $G$ has at most $40s = 40((\ell+1)n) \leq 40((\lceil 12 \log n \rceil +3)n)=O(n\log n)$ vertices. Hence, an $2^{o(n/\log n)}$ algorithm solving \mspd\ or \mstd\ implies by the reduction a $2^{o((n \log n)/(\log n - \log \log n))}=2^{o(n)}$ algorithm for \textsc{String 3-Groups}, which violates the ETH by Theorem~\ref{theorem:groups}.

\section{Conclusions}
In this paper, we showed that the time needed for the MSTD and MSPD problems, for fixed $k$, is $2^{\Theta(n/\log n)}$. For our lower bound,
we assume the ETH to hold, and need that $k\geq 39$. We expect that with more intricate constructions (see e.g., the gadgets used
in the proofs in \cite{DereniowskiKZ15}), the value of $39$ can be brought down; it is interesting to see if the lower bound
still holds for small values of $k$, e.g., $k=5$.

The intermediate results in our lower bound proof have independent interest, e.g.,
we conjecture that the {\sc Strings 3-Groups} lower bound (Theorem~\ref{theorem:groups})
can be used to show a similar lower bound for {\sc Intervalizing 6-Colored Graphs}, cf. \cite{BodlaendervR11}, and it was recently used by Bodlaender et al.~\cite{BodlaenderNZ15}
to obtain lower bounds for a collection of graph embedding problems (including {\sc Subgraph
Isomorphism} and {\sc Graph Minor} for graphs of pathwidth two or three).

The upper bound technique is also of independent interest and is likely to have more
applications. The central idea can be characterized as follows: where a standard
dynamic programming algorithm uses a canonical form of partial solutions, we add
a second level of canonization, by using graph isomorphism to find canonical forms of
the first level of canonical forms.
Thus, our results form a nice example of a technique which we would like to call
{\em supercanonization}.


%

It is also interesting to explore for which other problems the optimal time bound under the exponential time hypothesis is $2^{\Theta(n / \log n)}$. Bodlaender and Fomin~\cite{BodlaenderF05}
considered minimum cost versions of tree decompositions. For a function $f$, the $f$-cost of a tree decomposition is the sum over all bags $X$ of
$f(|X|)$. Consider the following problems, for some (sufficiently fast computable) function $f$: given a graph $G$ and integer $k$,
what is the minimum $f$-cost of a tree (or path) decomposition of width at most $k$. Our algorithms can be easily adapted for this problem, i.e.,
for fixed $k$, one can find a minimum $f$-cost tree (or path) decomposition of $G$ of width $k$ (if existing) in $2{O(n/\log n)}$ time, plus the
time to compute $f(0), f(1), \ldots, f(k+1)$. It is interesting to explore when this is optimal. We conjecture that the lower bound
proof can be adapted when $f$ is linear or sublinear. 
Bodlaender and Fomin~\cite{BodlaenderF05} have shown that for functions $f$ that fulfil for all $i$:
$f(i+1)\geq 2f(i)$, there is always a minimum cost triangulation that is a minimal triangulation;
such functions are called {\em fast}. This can be used to show that the problem of finding
a minimum $f$-cost tree decomposition of width at most $k$ (if existing), parameterized by $k$ belongs to XP, when $f$ is
fast and $f(0), \ldots, f(k+1)$ can be computed efficiently. This technique does not seem to apply in the case of path decompositions.

Another possible extension of our results is the case that $k$ is not constant. Using techniques from \cite{BodlaenderNZ15},
one can show that for $H$-minor free graphs (for fixed $H$), the MSPD and MSTD problems can still be solved
in $2^{O(n/\log n)}$ time. The technical details are a more or less straightforward combination of the techniques
in this paper and in \cite{BodlaenderNZ15}.

\paragraph{Acknowledgements.} We thank the anonymous referees for their detailed and helpful comments.

\bibliographystyle{abbrv}
\bibliography{definitions,papers}

\end{document}